# The New Ephemeris and Light Curve Analysis of V870 Ara by the Ground-Based and TESS Data

Atila Poro[1], Mark G. Blackford[2], Fatemeh Davoudi[1], Amirreza Mohandes[1], Mohammad Madani[1], Samaneh Rezaei[1], Elnaz Bozorgzadeh[1]

[1]The International Occultation Timing Association Middle East section, Iran, info@iota-me.com
[2]Variable Stars South (VSS), Congarinni Observatory, Congarinni, NSW, 2447, Australia

**ABSTRACT**
New CCD photometric observations and their investigation of the W UMa-type binary, V870 Ara, are presented. Light curves of the system were taken through $BVI$ filters from the Congarinni Observatory in Australia. The new ephemeris is calculated based on seven new determined minimum times, together with the TESS data and others compiled from the literature. Photometric solutions determined by the Wilson-Devinney (W-D) code are combined with the Monte Carlo simulation to determine the adjustable parameters' uncertainties. These solutions suggest that V870 Ara is a contact binary system with a mass ratio of 0.082, a fillout factor of 96±4 percent, and an inclination of 73.60±0.64 degrees. The absolute parameters of V870 Ara were determined by combining the Gaia EDR3 parallax and photometric elements.

## 1. INTRODUCTION

The contact binary system V870 Ara is located in the southern constellation Ara and has a magnitude of $V = 8.96$ and an orbital period of 0.399722 day (Eker et al. 2009, Szalai et al. 2007). V870 Ara is one of the systems discovered by the Hipparcos space-based telescope. Kazarovets et al. (1999) put this star in the Hipparcos variable stars' catalog and suggested that V870 Ara may be a Delta Scuti variable star. Selam (2004) analyzed the light curve of this system for the first time, calculated a mass ratio of $q = 0.25$, and put this variable star in the W-subtypes of the W UMa contact binary type. Szalai et al. (2007) spectroscopically measured $q = 0.082 \pm 0.030$ and computed this system's distance to be 112.5 parsecs. Szalai et al. (2007) and Pribulla & Rucinski (2006) studied the period change of five systems, including the V870 Ara, to identify the third body, and they could not find signs of multiplicity in the V870 Ara. Ulaş et al. (2012) classified the V870 Ara as a late-type contact binary system, whereas Hu et al. (2018) classified it as a deep-contact binary system.

In this study, the multi-color CCD light curves in $B$, $V$, and $I$ bands, along with photometric data obtained from the TESS space telescope, are presented. We extracted the minimum times based on the Monte Carlo Markov Chain (MCMC) method and then determined a new ephemeris for this binary system. The light curve solution with the W-D code combined with the Monte Carlo (MC) simulation was performed. Absolute parameters of the system were derived.

## 2. OBSERVATION AND DATA REDUCTION

The new photometric observations of V870 Ara in June 2020 were made using $BVI$ standard filters. These observations were carried out with the Orian ED80T CF refractor telescope at the Congarinni Observatory, located in Australia (152° 52´ E and 30°44´ S). A total of 1480 photos in the filters were taken by a CCD Atik One 6.0. This CCD has $2750 \times 2200$ pixels, 4.54-micron square, and we used $1 \times 1$ binning for observations. The temperature of the CCD is set at -10°C. In these observations, we selected a comparison star and a check star close to the V870 Ara. Aspects like apparent magnitude, spectral category, and distance from V870 Ara were taken into account when choosing the comparison and the check stars. The characteristics of all these stars are shown in Table 1, and the field of view is shown in Figure 1. MaxIm DL software was used to apply dark, bias, and flat-field corrections and perform aperture photometry.

Table 1. Characteristics of the binary system, comparison star, and check star (from Simbad[1]).

| Type | Name | RA. (J2000) | DEC. (J2000) | $V$ (mag.) | Sp. Type |
|---|---|---|---|---|---|
| Variable | V870 Ara | 18 08 22.6745 | -56 46 01.8007 | 8.96 | F7/G0 |
| Comparison | TYC 8751-1567-1 | 18 11 29.2004 | -56 36 55.7000 | 9.88 | G5 |
| Check | TYC 8751-1647-1 | 18 08 20.7684 | -56 45 30.4884 | 9.84 | G5 |

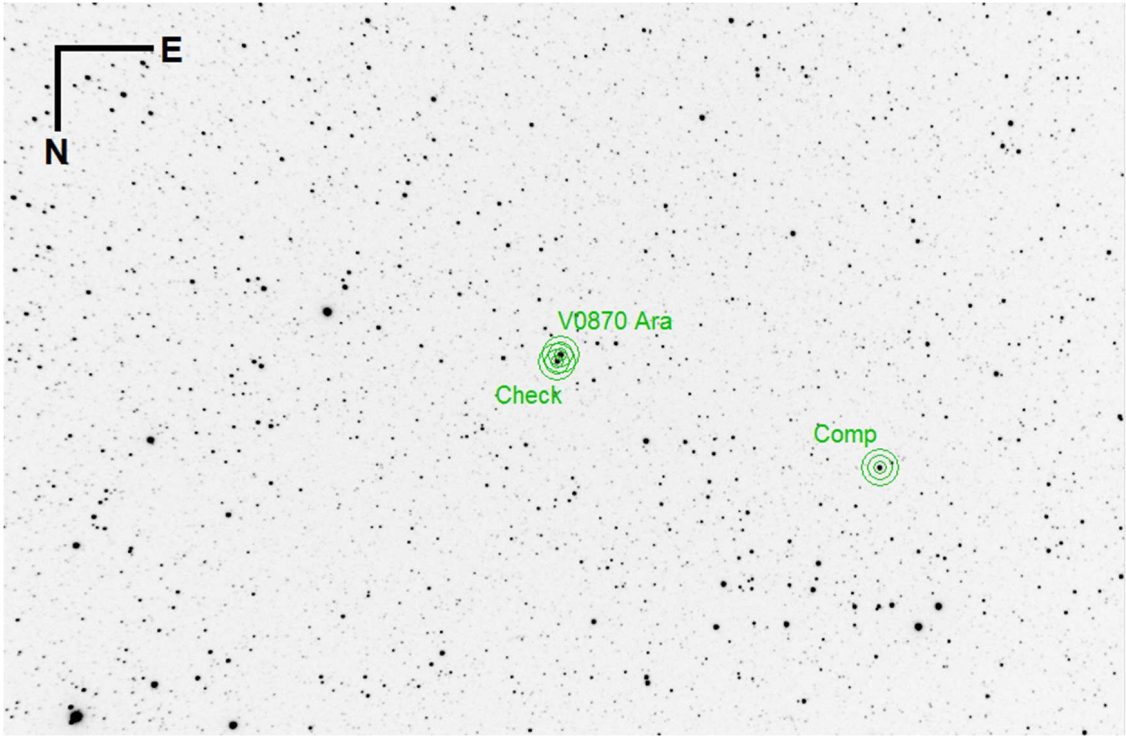

**Figure 1.** The position of V870 Ara, TYC 8751-1567-1 (comparison star), and TYC 8751-1647-1 (check star). The field of view is 90 by 72 arcminutes. The circles are only to indicate the position of each star with much larger rings. The aperture radius used for photometry was 8 pixels, gap width 5 pixels and annulus thickness 5 pixels.

The Transiting Exoplanet Survey Satellite (TESS) is NASA's two-year mission to search for exoplanets and variable stars and it was launched in 2018. TESS observed V870 Ara (TIC 118271395) from 19 June to 18 July, 2019, and the data is available at the Mikulski Space Telescope Archive (MAST). We extracted TESS style curves detrended by the TESS Science Processing Operations Center (SPOC) pipeline (Jenkins et al. 2016) from the MAST using the LightKurve code[2]. It was observed in sector 13 by Camera 2 and CCD 4 at a 120 second cadence. The data normalized by the AstroImageJ (AIJ) software (Collins et al. 2017).

**3. NEW EPHEMERIS**

To extract the minimum times and their uncertainties, we employed the Monte Carlo Markov Chain (MCMC) approach through fitting the models to the light curves based on Gaussian distributions (Poro et al. 2020). The PyMC3 package was used to implement the code for this part of the study (Salvatier et al. 2016). According to this, we found four primary and three secondary minima in our observed light curves. Using the same method, we also extracted 135 minimum times from the TESS data. We used a Python code based on the lightKurve package to extract SAP light curves from calibrated TESS data files obtained from the Mikulski Space Telescope Archive (MAST) (Lightkurve Collaboration et al. 2018). All the extracted minima times from our observation and TESS data with the collected minima times from the literature are listed in Table 2. All times of minimum are expressed in Barycentric Julian Date in Barycentric Dynamical Time (BJD$_{TDB}$) are in column 1, their uncertainties appear in column 2, epochs of these minima times in column 3, O-C values in column 4, and the references to minima times in the last column.

---
[1] http://simbad.u-strasbg.fr/simbad/
[2] https://docs.lightkurve.org/

We used the following light elements as the reference ephemeris (Szalai 2007) for computing epoch and initial O-C values in Table 2,

$$Min.I(BJD_{TDB}) = 2453185.13865(3) + 0.39972200(2) \times E. \quad (1)$$

**Table 2.** Times of minima of V870 Ara[3].

| Min. (BJD$_{TDB}$) | Error | Epoch | O-C (day) | Reference | Min. (BJD$_{TDB}$) | Error | Epoch | O-C (day) | Reference |
|---|---|---|---|---|---|---|---|---|---|
| 2453185.13865 | 0.00003 | 0 | 0 | Szalai 2007 | 2458669.52324 | 0.00014 | 13720.5 | -0.0011 | TESS |
| 2453195.13215 | 0.00003 | 25 | 0.0004 | Szalai 2007 | 2458669.72326 | 0.00011 | 13721 | -0.0010 | TESS |
| 2453196.13325 | 0.00003 | 27.5 | 0.0022 | Szalai 2007 | 2458669.92308 | 0.00013 | 13721.5 | -0.0010 | TESS |
| 2458654.13301 | 0.00007 | 13682 | -0.0020 | TESS | 2458670.12200 | 0.00012 | 13722 | -0.0019 | TESS |
| 2458654.53294 | 0.00011 | 13683 | -0.0018 | TESS | 2458670.32288 | 0.00013 | 13722.5 | -0.0009 | TESS |
| 2458654.53275 | 0.00011 | 13683 | -0.0020 | TESS | 2458670.52261 | 0.00010 | 13723 | -0.0010 | TESS |
| 2458654.73291 | 0.00014 | 13683.5 | -0.0017 | TESS | 2458670.72266 | 0.00012 | 13723.5 | -0.0009 | TESS |
| 2458654.93253 | 0.00010 | 13684 | -0.0020 | TESS | 2458670.92232 | 0.00009 | 13724 | -0.0011 | TESS |
| 2458655.13275 | 0.00013 | 13684.5 | -0.0016 | TESS | 2458671.12228 | 0.00012 | 13724.5 | -0.0010 | TESS |
| 2458655.33239 | 0.00009 | 13685 | -0.0018 | TESS | 2458671.32220 | 0.00009 | 13725 | -0.0009 | TESS |
| 2458655.53257 | 0.00013 | 13685.5 | -0.0015 | TESS | 2458671.52192 | 0.00012 | 13725.5 | -0.0010 | TESS |
| 2458655.73213 | 0.00009 | 13686 | -0.0018 | TESS | 2458671.72197 | 0.00010 | 13726 | -0.0009 | TESS |
| 2458655.93237 | 0.00013 | 13686.5 | -0.0014 | TESS | 2458671.92165 | 0.00012 | 13726.5 | -0.0010 | TESS |
| 2458656.13197 | 0.00008 | 13687 | -0.0017 | TESS | 2458672.12135 | 0.00011 | 13727 | -0.0012 | TESS |
| 2458656.33211 | 0.00013 | 13687.5 | -0.0014 | TESS | 2458672.32138 | 0.00016 | 13727.5 | -0.0010 | TESS |
| 2458656.53179 | 0.00008 | 13688 | -0.0016 | TESS | 2458672.52056 | 0.00014 | 13728 | -0.0017 | TESS |
| 2458656.73205 | 0.00013 | 13688.5 | -0.0012 | TESS | 2458672.72095 | 0.00015 | 13728.5 | -0.0012 | TESS |
| 2458656.93169 | 0.00010 | 13689 | -0.0014 | TESS | 2458672.92040 | 0.00011 | 13729 | -0.0016 | TESS |
| 2458657.13105 | 0.00014 | 13689.5 | -0.0019 | TESS | 2458673.12083 | 0.00014 | 13729.5 | -0.0010 | TESS |
| 2458657.33159 | 0.00010 | 13690 | -0.0012 | TESS | 2458673.32019 | 0.00010 | 13730 | -0.0015 | TESS |
| 2458657.53068 | 0.00017 | 13690.5 | -0.0020 | TESS | 2458673.52050 | 0.00016 | 13730.5 | -0.0011 | TESS |
| 2458657.73159 | 0.00015 | 13691 | -0.0010 | TESS | 2458673.71996 | 0.00009 | 13731 | -0.0015 | TESS |
| 2458657.93132 | 0.00015 | 13691.5 | -0.0011 | TESS | 2458673.92036 | 0.00013 | 13731.5 | -0.0009 | TESS |
| 2458658.13143 | 0.00012 | 13692 | -0.0008 | TESS | 2458674.11977 | 0.00009 | 13732 | -0.0014 | TESS |
| 2458658.33008 | 0.00015 | 13692.5 | -0.0021 | TESS | 2458674.31996 | 0.00012 | 13732.5 | -0.0011 | TESS |
| 2458658.53068 | 0.00010 | 13693 | -0.0013 | TESS | 2458674.51884 | 0.00009 | 13733 | -0.0020 | TESS |
| 2458658.73077 | 0.00013 | 13693.5 | -0.0011 | TESS | 2458674.71978 | 0.00012 | 13733.5 | -0.0010 | TESS |
| 2458658.92990 | 0.00008 | 13694 | -0.0018 | TESS | 2458674.91918 | 0.00009 | 13734 | -0.0014 | TESS |
| 2458659.13009 | 0.00012 | 13694.5 | -0.0015 | TESS | 2458675.11952 | 0.00012 | 13734.5 | -0.0009 | TESS |
| 2458659.32994 | 0.00009 | 13695 | -0.0015 | TESS | 2458675.31809 | 0.00009 | 13735 | -0.0022 | TESS |
| 2458659.52971 | 0.00012 | 13695.5 | -0.0016 | TESS | 2458675.31909 | 0.00009 | 13735 | -0.0012 | TESS |
| 2458659.72964 | 0.00009 | 13696 | -0.0015 | TESS | 2458675.71785 | 0.00009 | 13736 | -0.0022 | TESS |
| 2458659.93006 | 0.00012 | 13696.5 | -0.0010 | TESS | 2458675.91889 | 0.00012 | 13736.5 | -0.0010 | TESS |
| 2458660.13001 | 0.00008 | 13697 | -0.0009 | TESS | 2458676.11885 | 0.00009 | 13737 | -0.0009 | TESS |
| 2458660.32862 | 0.00012 | 13697.5 | -0.0021 | TESS | 2458676.31795 | 0.00013 | 13737.5 | -0.0017 | TESS |
| 2458660.52971 | 0.00007 | 13698 | -0.0009 | TESS | 2458676.51852 | 0.00010 | 13738 | -0.0010 | TESS |
| 2458660.72839 | 0.00012 | 13698.5 | -0.0021 | TESS | 2458676.71789 | 0.00012 | 13738.5 | -0.0015 | TESS |
| 2458660.92899 | 0.00011 | 13699 | -0.0013 | TESS | 2458676.91744 | 0.00009 | 13739 | -0.0018 | TESS |
| 2458661.12883 | 0.00014 | 13699.5 | -0.0014 | TESS | 2458677.11781 | 0.00012 | 13739.5 | -0.0013 | TESS |
| 2458661.32794 | 0.00009 | 13700 | -0.0021 | TESS | 2458677.31726 | 0.00008 | 13740 | -0.0017 | TESS |
| 2458661.52769 | 0.00013 | 13700.5 | -0.0022 | TESS | 2458677.51663 | 0.00012 | 13740.5 | -0.0022 | TESS |
| 2458661.72860 | 0.00010 | 13701 | -0.0012 | TESS | 2458677.71703 | 0.00008 | 13741 | -0.0016 | TESS |
| 2458661.92751 | 0.00013 | 13701.5 | -0.0021 | TESS | 2458677.91737 | 0.00012 | 13741.5 | -0.0011 | TESS |
| 2458662.12847 | 0.00008 | 13702 | -0.0010 | TESS | 2458678.11692 | 0.00008 | 13742 | -0.0015 | TESS |
| 2458662.32739 | 0.00012 | 13702.5 | -0.0020 | TESS | 2458678.31603 | 0.00011 | 13742.5 | -0.0022 | TESS |
| 2458662.52812 | 0.00008 | 13703 | -0.0011 | TESS | 2458678.51662 | 0.00008 | 13743 | -0.0015 | TESS |
| 2458662.72702 | 0.00012 | 13703.5 | -0.0021 | TESS | 2458678.71614 | 0.00011 | 13743.5 | -0.0018 | TESS |

---

[3] We ignored the literature's minimum times of three and less than three decimal places to improve accuracy.

| | | | | | | | | | |
|---|---|---|---|---|---|---|---|---|---|
| 2458662.92796 | 0.00008 | 13704 | -0.0010 | TESS | 2458678.91643 | 0.00010 | 13744 | -0.0014 | TESS |
| 2458663.12671 | 0.00013 | 13704.5 | -0.0021 | TESS | 2458679.11688 | 0.00012 | 13744.5 | -0.0008 | TESS |
| 2458663.32703 | 0.00008 | 13705 | -0.0016 | TESS | 2458679.31619 | 0.00008 | 13745 | -0.0013 | TESS |
| 2458663.52658 | 0.00013 | 13705.5 | -0.0019 | TESS | 2458679.51552 | 0.00013 | 13745.5 | -0.0019 | TESS |
| 2458663.72703 | 0.00009 | 13706 | -0.0014 | TESS | 2458679.71603 | 0.00009 | 13746 | -0.0012 | TESS |
| 2458663.92646 | 0.00013 | 13706.5 | -0.0018 | TESS | 2458679.91540 | 0.00012 | 13746.5 | -0.0017 | TESS |
| 2458664.12651 | 0.00012 | 13707 | -0.0016 | TESS | 2458680.11570 | 0.00008 | 13747 | -0.0013 | TESS |
| 2458664.32607 | 0.00015 | 13707.5 | -0.0019 | TESS | 2458680.31535 | 0.00012 | 13747.5 | -0.0015 | TESS |
| 2458664.52609 | 0.00011 | 13708 | -0.0017 | TESS | 2458680.51461 | 0.00007 | 13748 | -0.0021 | TESS |
| 2458664.72592 | 0.00014 | 13708.5 | -0.0018 | TESS | 2458680.71526 | 0.00011 | 13748.5 | -0.0013 | TESS |
| 2458664.92660 | 0.00011 | 13709 | -0.0009 | TESS | 2458680.91530 | 0.00006 | 13749 | -0.0011 | TESS |
| 2458665.12564 | 0.00014 | 13709.5 | -0.0018 | TESS | 2458681.11414 | 0.00011 | 13749.5 | -0.0021 | TESS |
| 2458665.52547 | 0.00014 | 13710.5 | -0.0017 | TESS | 2458681.31503 | 0.00007 | 13750 | -0.0011 | TESS |
| 2458665.72607 | 0.00010 | 13711 | -0.0009 | TESS | 2458681.51388 | 0.00011 | 13750.5 | -0.0021 | TESS |
| 2458665.92524 | 0.00014 | 13711.5 | -0.0016 | TESS | 2458681.71483 | 0.00007 | 13751 | -0.0010 | TESS |
| 2458666.12541 | 0.00010 | 13712 | -0.0013 | TESS | 2458681.91367 | 0.00011 | 13751.5 | -0.0021 | TESS |
| 2458666.32495 | 0.00014 | 13712.5 | -0.0016 | TESS | 2458682.11440 | 0.00006 | 13752 | -0.0012 | TESS |
| 2458666.52506 | 0.00010 | 13713 | -0.0014 | TESS | 2458682.31381 | 0.00011 | 13752.5 | -0.0016 | TESS |
| 2458666.72474 | 0.00013 | 13713.5 | -0.0016 | TESS | 2459016.08119 | 0.00035 | 14587.5 | -0.0021 | This study |
| 2458666.92441 | 0.00009 | 13714 | -0.0017 | TESS | 2459016.28093 | 0.00038 | 14588 | -0.0023 | This study |
| 2458667.12463 | 0.00014 | 13714.5 | -0.0014 | TESS | 2459016.08149 | 0.00030 | 14587.5 | -0.0018 | This study |
| 2458667.32408 | 0.00009 | 13715 | -0.0018 | TESS | 2459016.28060 | 0.00036 | 14588 | -0.0026 | This study |
| 2458667.52439 | 0.00013 | 13715.5 | -0.0014 | TESS | 2459003.09020 | 0.00024 | 14555 | -0.0022 | This study |
| 2458668.92385 | 0.00012 | 13719 | -0.0009 | TESS | 2459016.08143 | 0.00036 | 14587.5 | -0.0019 | This study |
| 2458669.12343 | 0.00016 | 13719.5 | -0.0012 | TESS | 2459016.28052 | 0.00036 | 14588 | -0.0027 | This study |

We refined new ephemeris for this system using fitting a line on all minima times based on the MCMC method (10000 steps, 200 walkers, burn-in=100) as shown in Figure 2. We wrote this code using the emcee package in Python (Foreman-Mackey et al. 2013). The code gives changes in the light elements of the reference ephemeris as an output. We corrected reference ephemeris as

$$Min. I\ (BJD_{TDB}) = 2453185.14068(95) + 0.39972197(5) \times E\ [days] \quad (2)$$

where $E$ is the integer number of orbital cycles after the reference epoch.

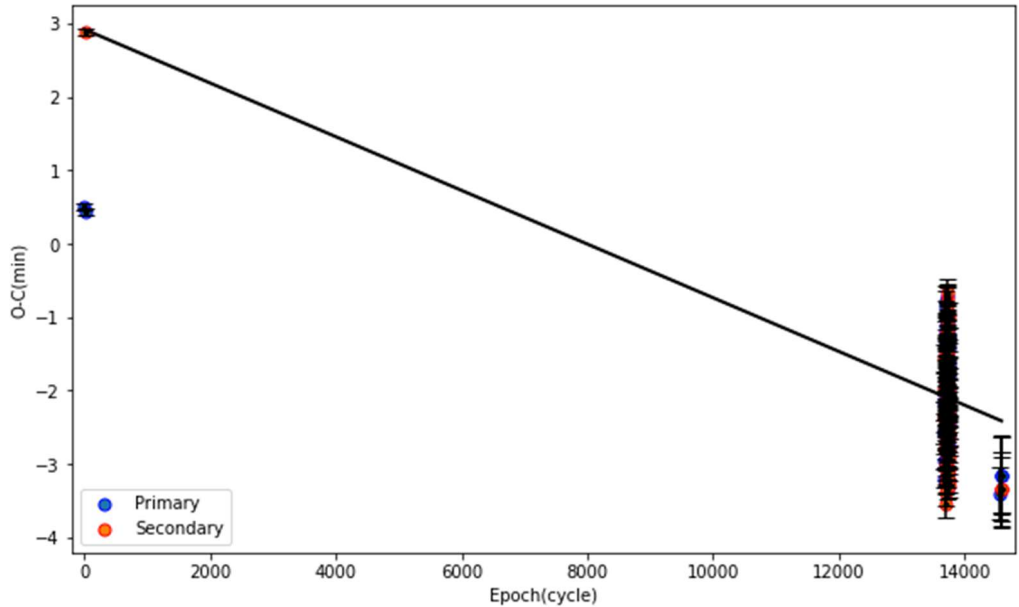

**Figure 2.** The O-C diagram of V870 Ara with the linear trend in the data using the MCMC method. The blue color corresponds to primary minima times, while the red color corresponds to secondary minima times.

**4. BINARY PARAMETER ANALYSIS**

Analysis of the light curves to determine physical parameters of the binary system was carried out using the Wilson & Devinney (1971) code (W-D) combined with the MC simulation to reduce degeneration in the solution space and determine the uncertainties of the adjustable parameters (Zola et al. 2004, 2010). We set the free parameters in the MC simulation and their ranges based on previous studies of this system and the light curves obtained from our observation (Table 3).

**Table 3.** Free parameters and search range in MC Simulations.

| Parameter | Value |
|---|---|
| $i$ (deg) | 50-90 |
| $T_2$ (K) | 5000-7000 |
| $\Omega_{1,2}$ | 1-10 |
| $l_1$ | 1-12 |
| $q$ | 0-5 |
| Phase shift | -0.03-0.03 |
| $X_{\text{Limb darkening}}$ | 0-1 |
| $Y_{\text{Limb darkening}}$ | 0-1 |
| co-latitude (deg) | 0-180 |
| Longitude (deg) | 0-360 |
| Spot radius (deg) | 1-90 |
| $T_{\text{spot}}/T_1$ | 0.7-1.1 |

The $q$-search analysis gives reliable mass ratios when full-eclipses are observed in the light curve (Kjurkchieva et al. 2019). At first, a range of fixed mass ratios from 0 to 5 was used to search. However, it only indicated minima in the range of 0.08 to 0.25 as shown in Figure 3. As a result, we found the minimum sum of the squared residuals of the W-D fit, $\sum(O-C)^2$ and $q = 0.821$. Szalai et al. (2007) determined $q = 0.82$ by a spectroscopic study on this system. Therefore, due to the close values, we fixed the 2007 study mass ratio for our final light curve solutions.

Since both the primary and the secondary stars were assumed to have convective gas, gravity darkening exponents ($g_1 = g_2$) and albedos ($A_1 = A_2$) were fixed to theoretical reference values for convective gas, which are 0.32 and 0.5, respectively (Lucy 1967). Although the amount of limb darkening coefficients is considered a free parameter, the final values were in good agreement with Van Hamme's tables (1993).

According to the Phase-Flux light curve, the primary minimum is deeper but the secondary minimum happened first. The temperature of the primary star was fixed in the Szalai et al. (2007) analysis, which is the same condition as the difference in minimum deepening and the position of each minimum. We used the Gaia DR2 temperature value (5760 K) to set the temperature of the primary component.

Based on our data and after the required calibrations (Høg et al. 2000), we calculated $(B-V)_{V870\,Ara} = 0^m.632$. Thus, the effective temperature of the primary component, $T_1$ was assumed as 5777 K (Flower 1996). This temperature value is a good approximation because we can compare it with that of the Gaia DR2 catalog. The temperature difference is consistent; this indicates that the accuracy of the observations is reliable.

The depths of the primary and secondary minima in the light curves are different, indicating that the two stars' temperatures are not the same. As a result, we used the "overcontact binary not in thermal contact" mode. The parameters and uncertainties obtained from the light curve solutions in the $BVI$ filters are presented in Table 4. Despite setting the range for the starpot model as free parameters (Table 3), there was need to add a hot-starspot due to the light curve solutions. Figures 4 shows our observations, TESS, and synthetic light curves.

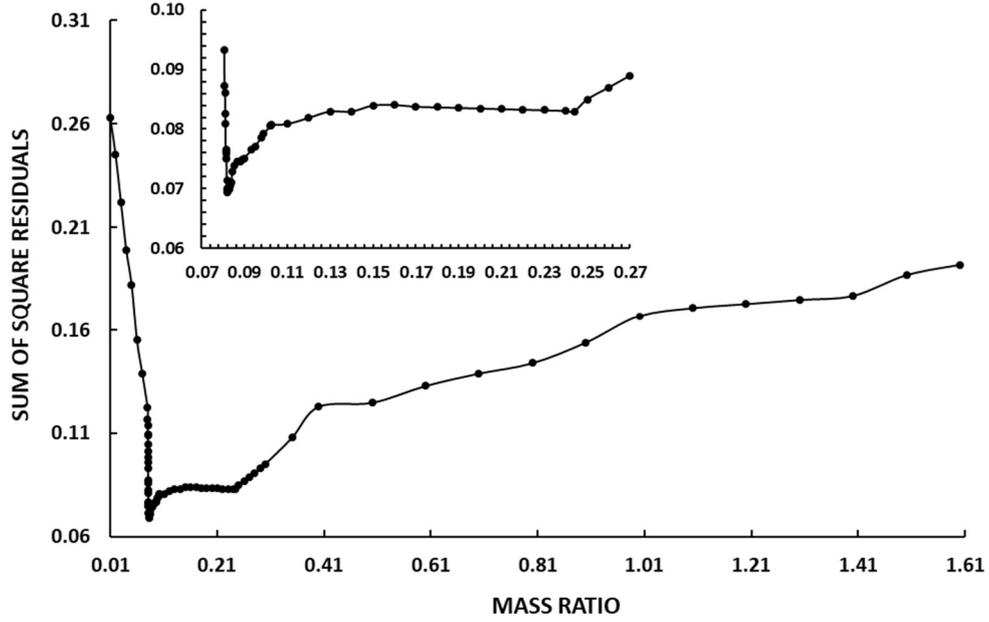

**Figure 3.** Sum of the squared residuals as a function of the mass ratio. The minimum part is zoomed.

Table 4. Photometric solutions of V870 Ara.

| Parameter | This study |
|---|---|
| $T_1$ (K) | 5760 |
| $T_2$ (K) | 5875(159) |
| $\Omega_1 = \Omega_2$ | 1.834(199) |
| $i$ (deg) | 73.60(64) |
| $q = M_2/M_1$ | 0.082 |
| $l_1/l_{tot}(BVI)$ | 0.863(55) |
| $l_2/l_{tot}(BVI)$ | 0.138(55) |
| $l_1/l_{tot}(TESS)$ | 0.863(5) |
| $l_2/l_{tot}(TESS)$ | 0.138(5) |
| $l_3$ | 0 |
| $A_1 = A_2$ | 0.50 |
| $g_1 = g_2$ | 0.32 |
| $f$ (%) | 96(4) |
| $r_1$(mean) | 0.605(4) |
| $r_2$(mean) | 0.233(4) |
| Colatitude$_{spot}$ (deg) | 98 |
| Longitude$_{spot}$ (deg) | 84 |
| Radius$_{spot}$ (deg) | 17 |
| $T_{spot}/T_{star}$ | 1.03(1) |
| Phase Shift | -0.006(3) |

Note: A hot-starspot is on the primary component.

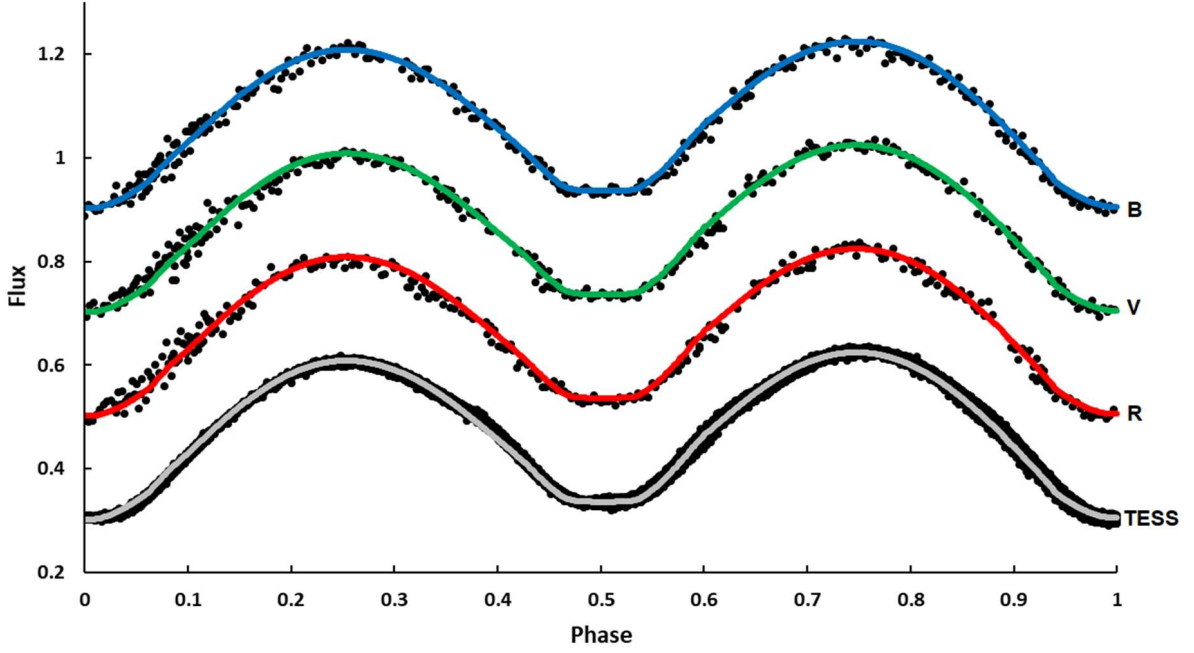

**Figure 4.** Light curves from our observations and TESS (black dots). The synthetic light curves obtained from light curve solutions in the $BVI$ filters and TESS are shown (top to bottom respectively); with respect to orbital phase, shifted arbitrarily in the relative flux.

We used the parallax from Gaia EDR3 and our photometry results for estimating absolute parameters of the binary system, and we calculated $d(pc) = 103.062 \pm 0.204$. According to our observation $V_{system} = 8^m.67 \pm 0.09$ and calculated $A_{dV} = 0.059 \pm 0.006$ (Schlafly & Finkbeiner 2011), the value of absolute magnitude, $M_{v(system)} = 3.546 \pm 0.079$, was estimated using an equation,

$$M_v = V - 5\log(d) + 5 - A_v \qquad (3)$$

As our light curve solutions conclude that $l_1/l_2 = 6.273$, we calculated $M_{v(1)} = 3.707 \pm 0.072$ and $M_{v(2)} = 5.700 \pm 0.036$. Also, with applying $(BC) = -0.022$ (Eker et. al 2020), $M_{bol(1)}$ and $M_{bol(2)}$ computed. We calculated $L$ and $R$ for each component of this system, using the well-known relations (Equations 4 and 5), respectively.

$$M_{bol(star)} - M_{bol(sun)} = -2.5\log\left(\frac{L_{star}}{L_{sun}}\right) \qquad (4)$$

$$L = 4\pi r^2 \sigma T^4 \qquad (5)$$

Equation $a = R/r$ gives us $a_1$ and $a_2$, from $R_1$ and $R_2$, and also, we reckoned the average of $a_1$ and $a_2$. Finally, we calculated the masses of the components, $M_1$ and $M_2$, by Kepler's third law and our $q = \left(\frac{M_2}{M_1}\right)$ value. The parameters obtained during this procedure are given in Table 5 compared to Szalai et al. (2007).

**Table 5.** Estimated absolute elements of V870 Ara along with the results of Szalai et al. (2007).

| Parameter | This study | | Szalai et al. (2007) | |
|---|---|---|---|---|
| | Primary | Secondary | Primary | Secondary |
| $Mass\ (M_\odot)$ | 1.546(54) | 0.127(37) | 1.503(11) | 0.123(2) |
| $Radius\ (R_\odot)$ | 1.64(6) | 0.63(5) | 1.67(1) | 0.61(1) |
| $Luminosity$ ($L_\odot$) | 2.64(17) | 0.42(1) | 2.96(30) | 0.50(1) |
| $M_{bol}$ (mag) | 3.69(7) | 5.68(4) | 3.54(10) | 5.48(10) |
| $log\ g$ (cgs) | 4.197(5) | 3.943(15) | 4.169(2) | 3.957(7) |
| $a$ ($R_\odot$) | 2.71(12) | | 2.43(13) | |

Figure 5 presents schematic representations of the two stars' relative shapes and sizes at four phases, showing their significant ellipsoidal distortion.

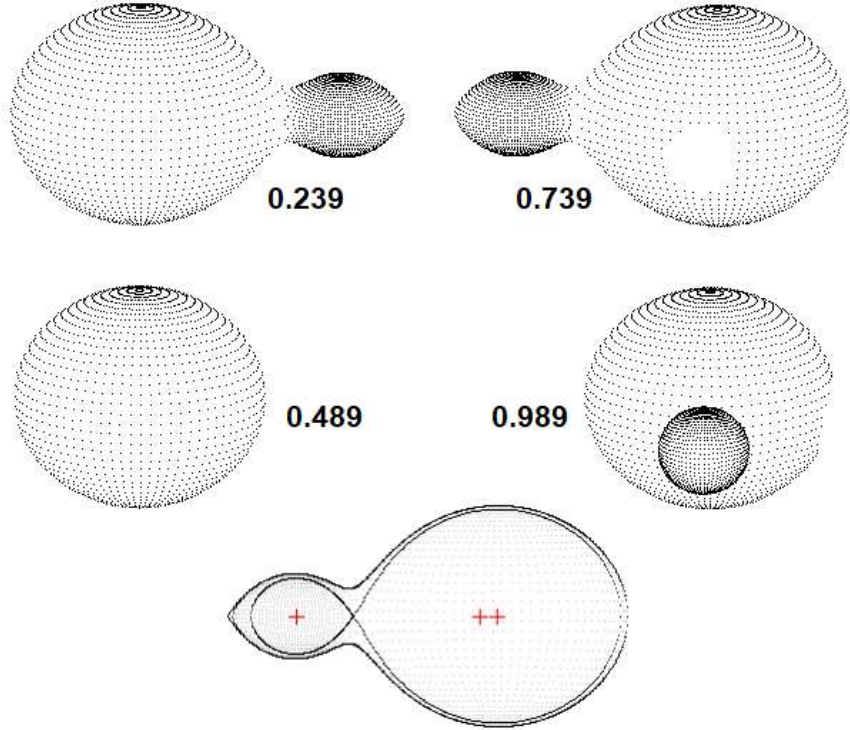

**Figure 5.** The geometrical structure of V870 Ara at the four phases, and the cross-sectional outline of the binary system at the phase of 0.739 with contact degree $f = 96\%$.

### 5. SUMMARY AND CONCLUSION

We obtained minimum times from new $BVI$ goround-based, and TESS observations of V870 Ara using Python code. We presented a new ephemeris considering the number of minimum times during observations (2004-2020) was low. The O-C diagram shows a decrease in the orbital period that probably originated from the accumulation of measurement errors in light elements of the reference ephemeris.

We set Gaia DR2's temperature value for the primary star, although $(B - V)$'s value was close to this temperature. We used a mass ratio value comes from Szalai et al. (2007) study ($q = 0.082$), however our $q$-search showed almost the same value. According to the W-D code and MC simulation, it is found that V870 Ara is a contact binary with a fillout factor of $f$=96±4%, and an inclination of $i$=73.60±0.64 degrees.

We estimated the system's absolute parameters using Gaia EDR3 parallax and light curve solutions from our observations. Based on this, we obtained $A_v$ and $BC$ values from valid references and used the orbital period, the magnitude in $V$ band, $l_{1,2}$, $r_{1,2(mean)}$, and mass ratio values from our calculations in this study. Therefore, the mass, radius, bolometric magnitude, and luminosity of the system were obtained. Table 5 shows that the absolute parameter values obtained in this investigation are comparable and consistent with those found in the Szalai et al. (2007) study.

According to the system's color index, the spectral type of V870 Ara is suggested as G2V from Allen's table (Cox 2000). The positions of the primary and secondary components of V870 Ara in the Mass-Radius ($M - R$) and Mass-Luminosity ($M - L$) diagrams on a $log$-scale are shown in Figure 6. These diagrams show the evolutionary status of V870 Ara.

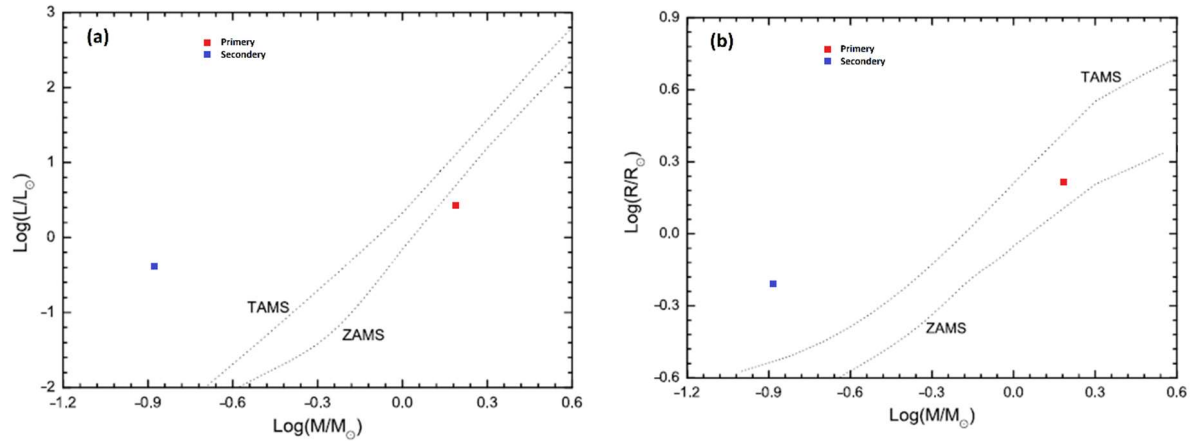

**Figure 6.** The $logM - logR$, and $logM - logL$ diagrams for V870 Ara from the absolute parameters.

The O'Connell effect (O'Connell 1951) can be recognized clearly in the light curves from the TESS and our observations. The most appropriate suggestion for this effect is the presence of starspot(s) induced by the components' magnetic activities (Sriram et al. 2017). The light curves of V870 Ara indicate $Max\ II$ is brighter than $Max\ I$ and asymmetry in maxima or unequal minima is clearly visible. Due to the presence of this asymmetry we used a stellar spot model during the light curve solutions. We found that assuming a hot-starspot model on the massive primary component results in an acceptable solution for all light curves.

Selam (2004) investigated the light curve solution for this binary system, reporting a mass ratio of 0.25 and a fillout factor of 70%. This system was also studied by Szalai et al. (2007), who determined $q$=0.082 and $f$=96.4%. The orbital inclination was estimated to be 70 degrees in both studies. We fixed the mass ratio value based on the Szalai et al. (2007) study. In comparison to the previous two studies, the orbital inclination value has increased slightly. As shown by the first and second minimums of the light curves, the temperature difference between the components has decreased in our study compared to the 2007 study.
We can conclude that V870 Ara is a contact and A-type W UMa system based on the low mass ratio, large fillout factor, and temperature difference between the components. Given that the amount of evidence for a low mass ratio, we expect it will become a deeper contact binary based on system evolution.

**Acknowledgments**